\documentclass[preprints,article,accept,moreauthors,pdftex]{Definitions/mdpi} 

\firstpage{1} 
\pubvolume{1}
\issuenum{1}
\articlenumber{0}
\pubyear{2021}
\copyrightyear{2020}
\datereceived{} 
\dateaccepted{} 
\datepublished{} 
\hreflink{https://doi.org/} 

\usepackage{subfig}
\usepackage{subfloat}
\usepackage{siunitx}
\usepackage{color}
\usepackage{colortbl} 
\usepackage{xcolor}
\newcommand{\red}[1]{\textcolor[rgb]{0.5,0,0}{#1}} 

\Title{Multi-Spectral Multi-Image Super-Resolution of Sentinel-2 with Radiometric Consistency Losses and Its Effect on Building Delineation}

\TitleCitation{Multi-Spectral Multi-Image Super-Resolution of Sentinel-2 with Radiometric Consistency Losses and Its Effect on Building Delineation}

\Author{Muhammed T. Razzak$^{1,}$*, Gonzalo Mateo-Garcia $^{2}$ Luis G{\'o}mez-Chova $^{2}$, Yarin Gal $^{1}$ and Freddie Kalaitzis $^{1,}$*}

\AuthorNames{Muhammed T. Razzak, Gonzalo Mateo-Garc{\'i}a, Luis G{\'o}mez-Chova, Yarin Gal, and Freddie Kalaitzis}

\AuthorCitation{Razzak, M.T.; Mateo-Garc{\'i}a, G.; G{\'o}mez-Chova, L.; Gal, Y.; Kalaitzis, F.}

\address{%
$^{1}$ \quad Oxford Applied and Theoretical Machine Learning Group, Department of Computer Science, University of Oxford; \{muhammed.razzak,freddie.kalaitzis, yarin\}@cs.ox.ac.uk\\
$^{2}$ \quad Image Processing Laboratory, University of Valencia; \{gonzalo.mateo-garcia, luis.gomez-chova\}@uv.es}

\corres{Correspondence: {muhammed.razzak,freddie.kalaitzis}@cs.ox.ac.uk}

\abstract{High resolution remote sensing imagery is used in broad range of tasks, including detection and classification of objects. High-resolution imagery is however expensive, while lower resolution imagery is often freely available and can be used by the public for range of social good applications. To that end, we curate a multi-spectral multi-image super-resolution dataset, using PlanetScope imagery from the SpaceNet 7 challenge as the high resolution reference and multiple Sentinel-2 revisits of the same imagery as the low-resolution imagery. We present the first results of applying multi-image super-resolution (MISR) to multi-spectral remote sensing imagery. We, additionally, introduce a radiometric consistency module into MISR model the to preserve the high radiometric resolution of the Sentinel-2 sensor. We show that MISR is superior to single-image super-resolution and other baselines on a range of image fidelity metrics. Furthermore, we conduct the first assessment of the utility of multi-image super-resolution on building delineation, showing that utilising multiple images results in better performance in these downstream tasks.}

\keyword{super-resolution; multi-image super-resolution; Sentinel 2; segmentation; detection; building detection} 

\begin{document}
\section{Introduction}
Generative Deep Learning has sparked a new wave of Super-Resolution (SR) algorithms that enhance the spatial resolution of images with impressive aesthetic results~\cite{wang2020deep}. Although the perceptual quality of those images is high, it is well-known that some of these SR models introduce artefacts into the SR image that are not present in real images~\cite{bhadra2020hallucinations}. In addition, most of these models do not enforce physically-based consistency between the super-resolved image and its low-resolution counterpart~\cite{bahat2020explorable}. This limits the applicability of SR models to domains such as remote sensing where the safety and consistency are critical, e.g. for scientific instrumentation and decision making.

Super-resolution models are divided in \textit{single-image} super-resolution (SISR) and \textit{multi-image} super-resolution (MISR) (aka.\textit{multi-frame} or \textit{multi-temporal} super-resolution). The former uses as input only one low-resolution image while the later takes several low-resolution images from the same scene. MISR seeks to further constrain the ill-posed problem of SR by conditioning on several low-res input images (aka. revisits). Therefore it is expected of MISR to produce better SR images, to be more robust and to produce fewer artifacts than SISR. In addition, MISR can be naturally applied in Earth observation since satellites often have frequent revisits of an area of interest. These multiple revisits can be fused with MISR to produce a super-resolved image. Despite its clear applicability, MISR has been scarcely applied in Remote Sensing and, as of yet, there are no studies that quantitatively compare MISR and SISR. Thus far, for remote sensing MISR has only been demonstrated only on RED and NIR bands of PROBA-V--- a tiny fraction of the Sentinel-2 operation spectrum \cite{deudon_highres-net_2020}. In addition, the practical utility of these super-resolved images for downstream tasks has been largely unexplored in real-world applications \cite{shermeyer2019effects}.

\begin{figure}
    \centering
    \includegraphics[width=\linewidth]{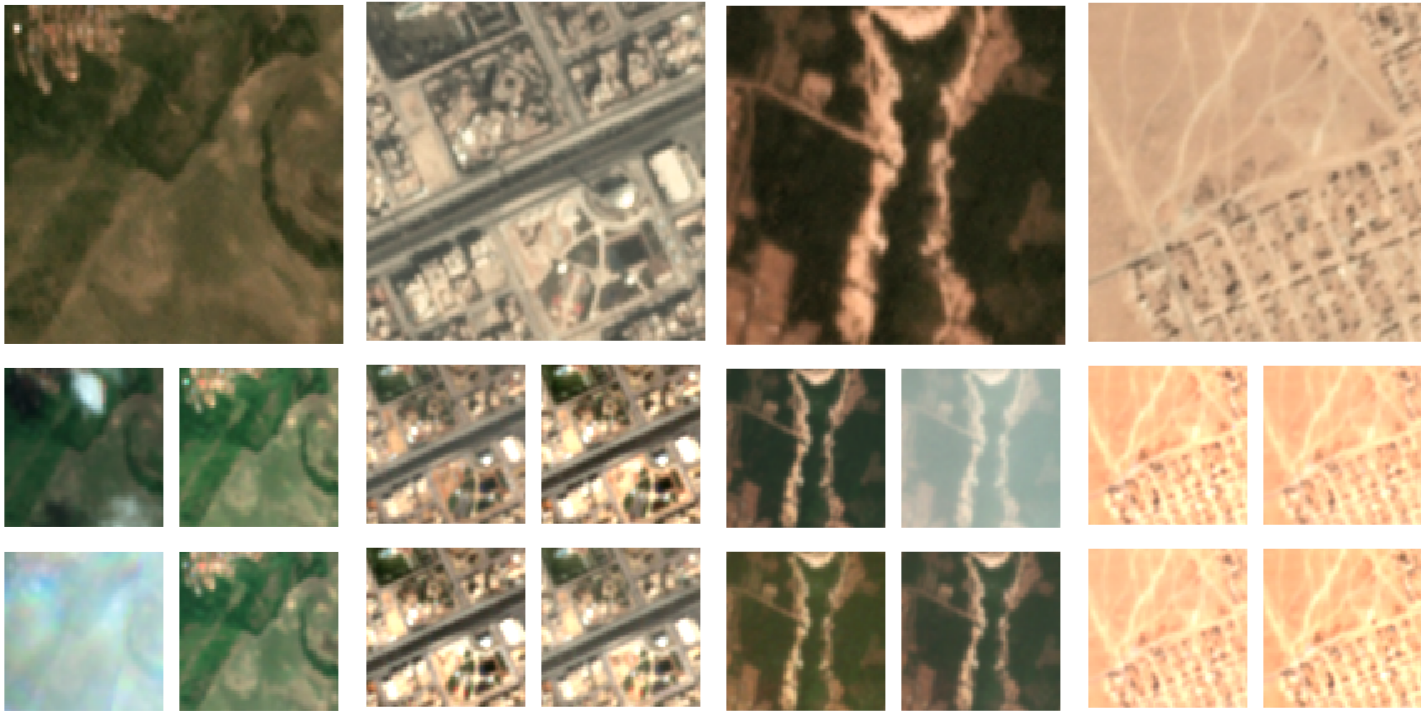}
    \caption{
        Different co-aligned retrievals from PlanetScope and Sentinel-2 from the SpaceNet 7 dataset. First row: PlanetScope RGB. Second and third rows: Sentinel-2 RGB revisits.
    }
    \label{fig:s2planet}
\end{figure}

In this work, we apply MISR to the multi-spectral multi-temporal satellite imagery from the European Space Agency's Copernicus Sentinel-2 (S2) archive, and study the downstream utility of the super-resolved images. In particular, we train super-resolution models (both SISR and MISR) on $~$\SI{10}{m} the RGB bands of Sentinel-2 images, using as reference co-registered \SI{4.77}{m} RGB PlanetScope images, with all imagery from within the same two-month period. This setting differs from the vast majority of previous remote sensing applications of SR, where low-res images are obtained by artificially downsampling the high-res counterpart \cite{shermeyer2019effects}. The main benefit of our setting is that the trained model can be applied on new S2 RGB images to enhance their nominal resolution to \SI{4.77}{m}, i.e. it provides out-of-sample SR results without requiring simultaneous and co-registered VHR images. We compared this model with SISR models on an independent test set showing better qualitative and quantitative performance in terms of PSNR (Peak Signal to Noise Ratio) and SSIM (Structural Similarity Index Measure), see sec.~\ref{sec:results}. In addition, we demonstrate that this resolution enhancement produce significant gains in tasks such as building delineation (sec.~\ref{sec:utilitymisr}).

One of the drawbacks of training super-resolution models when the low-res and high-res images come from different sensors, is the difference in the spectral characteristics of the sensing instruments and in the calibration of their output images. This poses the additional challenge of disentangling the SR task from the cross-instrument calibration task. Figure~\ref{fig:s2planet} shows several co-located Sentinel-2 and PlanetScope images (for each PlanetScope image, we show four S2 revisits within the same two-month period). We can see that there are differences in the colours of these images, in particular, Sentinel-2 images are brighter and with higher contrast and colours seem better defined. This is because Sentinel-2 images benefit from a higher radiometric and spectral resolution compared to PlanetScope, and both products undergo different atmospheric correction procedures to recover surface reflectance. In order to produce super-resolved images while preserving the spectra of the low-res Sentinel-2 imagery, we propose a radiometric consistency module to the super-resolution model, that helps to maintain the radiometric consistency of a super-resolved image with its low-res counterpart.


The contributions of our work are summarized as follows:
\begin{enumerate}
    \item We curate a new multi-temporal dataset from many revisits of Sentinel-2 imagery co-located with PlanetScope imagery, originally sourced from the SpaceNet-7 competition~\cite{van_etten_multi-temporal_2021}, which includes a geo-diverse set of scenes from around the globe.
    \item We demonstrate, for the first time, the multi-image super-resolution of RGB satellite imagery.
    \item We show that MISR is better at super-resolving Sentinel-2 RGB images compared to SISR, both quantitatively in terms of the image fedility metrics (PSNR and SSIM) and qualitatively.
    \item We demonstrate the downstream utility of super resolution, through the task of building semantic segmentation and instance segmentation.
    \item Finally, we propose a radiometric consistency module which can be added to any vanilla super-resolution model. We show that this module helps to maintain the radiometric consistency of Sentinel 2 while enhancing its spatial resolution, and we show several instances of good and bad consistencies.
\end{enumerate}
\section{Materials and Methods}
\subsection{Datasets}
\label{sec:datasets}

In order to learn a super-resolution model to improve the spatial resolution of S2, we need higher-resolution images to use as a reference. Since VHR (Very High Resolution) images (less than \SI{10}{m}) are not free, we restricted our search to pre-released publicly-available datasets of high-resolution images. Among those, we chose the recently launched Multi-temporal urban development SpaceNet dataset of PlanetScope images (also known as SpaceNet-7, see sec.~\ref{subsec:spacenet7})~\cite{van_etten_multi-temporal_2021}. With this dataset, we acquired co-located time series of Sentinel-2 images for each PlanetScope acquisition (sec.~\ref{subsec:sentinel2}). Subsection~\ref{subsec:analysissplits} has a brief analysis of the S2-Planet dataset as well as details about the different train-test splits that we used for the results.

\subsubsection{PlanetScope SpaceNet-7 dataset} \label{subsec:spacenet7}
\begin{figure}
    \centering
    \includegraphics[width=.85\linewidth]{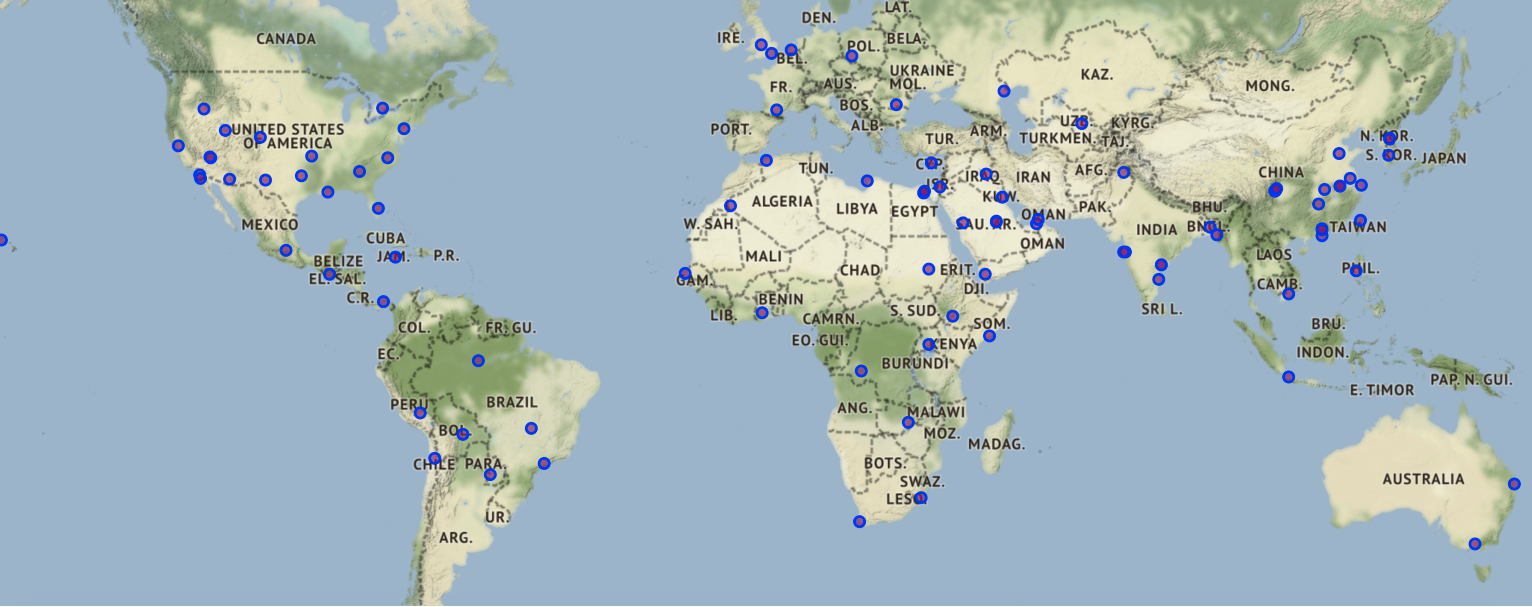}
    \caption{Location of SpaceNet-7 image time series. Figure taken from~\cite{van_etten_multi-temporal_2021}.}
    \label{fig:spacenet7Map}
\end{figure}
\begin{figure}
    \centering
    \includegraphics[width=.22\linewidth]{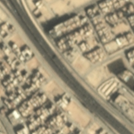}
    \includegraphics[width=.22\linewidth]{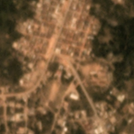}
    \includegraphics[width=.22\linewidth]{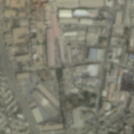}
    \includegraphics[width=.22\linewidth]{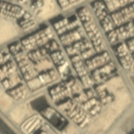}\\
    \includegraphics[width=.22\linewidth]{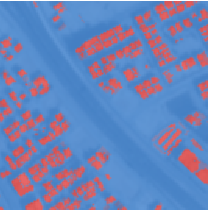}
    \includegraphics[width=.22\linewidth]{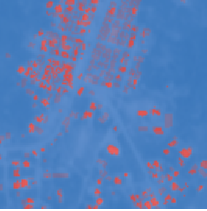}
    \includegraphics[width=.22\linewidth]{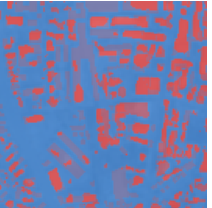}
    \includegraphics[width=.22\linewidth]{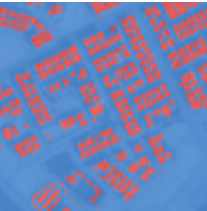}\\
    \caption{PlanetScope acquisitions (first row) and their corresponding building polygon masks (second row) from the SpaceNet-7 dataset.}
    \label{fig:footprints}
\end{figure}

SpaceNet-7 has monthly time series of PlanetScope images over a two-year time span period for approximately 100 different areas of interest (AOI) all over the world (see figure~\ref{fig:spacenet7Map}). Images are provided at \SI{4.77}{m} nominal\footnote{The resolution reported in the GeoTIFF metadata.} resolution with only three spectral channels (RGB). In addition, each of those images have manually derived polygon labels of building footprints. The challenge accompanying the release of this dataset consisted of delineating those buildings and identifying when new buildings were constructed on those areas. Figure~\ref{fig:footprints} shows some PlanetScope acquisitions with their corresponding building annotations over different locations. We can see that delineating the buildings on some of these scenes is quite challenging even at the resolution of PlanetScope.
In this study we restricted to images from December 2019 and January 2020 from the training set (building footprints in the test set have not been released yet). In total there are \red{46} different PlanetScope scenes for each month. 

\subsubsection{Sentinel-2 acquisitions} \label{subsec:sentinel2}

The Sentinel-2 mission consists of two satellites carrying the same multi-spectral optical sensor which acquires images on 13 different bands of the electromagnetic spectrum, from the visible to the short-wave infrared. The nominal spatial resolution of those images is different for each set of bands: 4 bands (visible and near infra-red) have \SI{10}{m} resolution; 6 bands in the very near infrared and short-wave infrared bands have \SI{20}{m} resolution; the remaining 3 bands are used mainly for atmospheric correction and they have a spatial resolution of \SI{60}{m}. Level 2A Sentinel-2 products, often referred as analysis ready data, consist of atmospherically corrected ortho-corrected 12-band images with bottom-of-atmosphere (BOA) calibrated reflectances. Providing good BOA calibrated images is one of the main goals of the Sentinel-2 mission, since an accurate radiometric and spectral calibration has large impacts on ocean (see e.g.~\cite{ruescas_machine_2018}) and vegetation products (e.g.~\cite{wolanin_estimating_2019,svendsen_joint_2018}). In section \ref{subsec:spectralconsistency}, we propose a SR model that seeks to maintain the spectral calibration of Sentinel-2.

Sentinel-2 images were downloaded from the ESA Open Access Hub. In order to obtain co-aligned time series of Sentinel-2 and PlanetScope images, we developed a custom pipeline which consists of the following steps:
\begin{enumerate}
    \item Download all Sentinel-2 level 2A products overlapping with each of the 46 PlanetScope scenes over December 2019 and January 2020. 
    \item Crop all Sentinel-2 images to the PlanetScope scene bounds.
    \item Reproject all bands of S2 to the coordinate reference system of PlanetScope products at \SI{10}{m} spatial resolution. 
    \item When more than one S2 product was found for the same date and scene, we mosaiced those images.
\end{enumerate}

Figure~\ref{fig:s2planet} shows different Planet (top) and Sentinel-2 (bottom) retrievals. As we can see, some of the Sentinel-2 images contain clouds. Sentinel-2 products have additional quality assessment bands that include cloud probabilities for each pixel. In this work, we used the SLC band to assess the presence of clouds; in particular, we used the value 9 which encodes "clouds high probability". This cloud indicator is used to inform the fusion model when merging different Sentinel-2 revisits.

\begin{figure}
    \centering
    \includegraphics[width=0.72\textwidth]{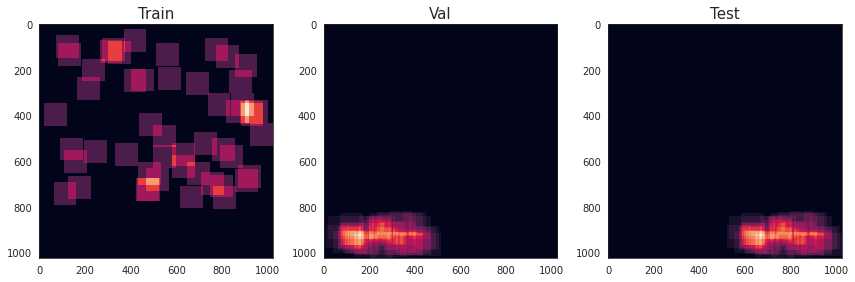}
    \caption{We utilised a within-scene split, which allocates the top 80\% of a scene as a source of training patches; the bottom-left and right 10\% for validation and testing patches respectively.
    }
    \label{fig:splits}
\end{figure}

\subsubsection{Dataset Analysis and training splits}\label{subsec:analysissplits}

The images come from \red{46} locations, with a geodiverse set of features including vegetation, bare earth (flats, hills, ridges), desert, urban, and agriculture infrastructure (see Appendix Table \ref{tab:AOI}). The number of revisits between December 2019 and January 2020 range from 5 to 13, but the percentage of usable revisits ($<50\%$ cloud coverage) ranges from 23\% to 100\%. Finally, we use a fixed partitioning of the scenes into training, validation and test datasets.

\subsubsection{Training, validation and test sets splits} 
\label{subsubsec:splits}

A well-thought split of the data into training and testing is critical to demonstrate the capacity of machine learning models to generalize. In remote sensing scenarios, extra-care must be taken to avoid train-test leakage due to spatial correlation. For instance, \cite{ploton_spatial_2020} recently showed that lack of consideration to spatial correlation lead to over-inflated results of ML models that monitored forest biomass. Our approach splits each scene is patches avoiding spatial overlap between patches in the different subsets; with this approach we seek to explore the performance of the models in ideal conditions when training and testing patches come from similar distributions. Figure \ref{fig:splits} shows the dataset partition for one scene.

\subsection{Metrics}
In this work, we look at two broad tasks: super-resolution and building delineation.

For super-resolution, the primary quantiative metrics of performance are the peak signal-to-noise ratio (PSNR) and the structural similarity index
(SSIM)~\cite{wang_image_2004} between the super-resolution image and the reference high-resolution image. Both measure the fidelity of the images compared, but with SSIM more focused on the structures contained in the images. Mathematically, PSNR is computed as follows:

\begin{equation}
PSNR = 10log \frac{I_{max}}{\mathrm{MSE}(HR , SR)}
\end{equation}
where $I_{max}$ is the maximum pixel value of the images (e.g. for 8-bit images this is 255), MSE is the mean square error between the high-resolution image ($HR$) and super-resolved image ($SR$).For the formula of the SSIM the reader is referred to the original work of \cite{wang_image_2004}.

\section{Multi-spectral multi-image super-resolution}

In the context of remote sensing, our work is the first to tackle the MISR problem in multiband images (RGB bands). Although the ultimate goal is multi-spectral MISR in all 13 bands of Sentinel-2, such a task is predicated on the coverage of the same bands by the high-resolution instrument, at least in the proposed supervised learning setting. Unfortunately, most VHR (very high resolution) products available are limited to R,G,B, and NIR (near-infrared) bands.

Back in traditional on-the-ground imaging, MISR has seen several successful applications on colour photos. Most notably, \cite{wronski2019handheld} achieved real-time on-board multi-image super-resolution in handheld Google Pixel cameras, by leveraging the user's hand motion jitter during a burst-frame shot. With intimate knowledge of the handheld camera's specifications (physical model), and no feature-learning involved, the authors could anticipate the amount of aliasing and phase shifting that images undergo after optical zoom-out and motion-jitter respectively. The fusion of burst-frames is ultimately a convex optimization problem (ADMM, see \cite{boyd2011distributed}). More recently, a deep learning approach to bust-frame MISR was proposed in \cite{bhat2021deep}, which is agnostic to camera specs.

\subsection{MISR in Earth Observation} \label{subsec:MISR_for_EO}

In handheld camera imaging, the problem of MISR is always learned with high-res reference images that comes from the same sensor. But in Earth Observation, the fact that any one instrument orbits at a fixed altitude, rules out any possibility of obtaining simultaneous low-res/high-res images pairs by the same sensor.  
Hence, this necessitates the use of a different higher resolution instrument, if MISR is to be learned in a supervised way. So unless the low-res and high-res pairs are acquired with different lens on-board the same satellite (e.g. PROBA-V, and a unique example at that, see \cite{mrtens2019superresolution}), then any ML-based MISR model must also learn to calibrate its output to the spectra and noise of the high-res instrument. In addition, for remote sensing the multiple images are temporally spaced over days and weeks, rather than over a few seconds with handheld imagery.

Recently, \cite{molini_deepsum_2020} and \cite{deudon_highres-net_2020} tackled the MISR problem in Earth Observation, in single-band imagery, and on the rather controlled use-case of PROBA-V that was the topic of ESA's MISR competition ending Q2-2019. In particular, \cite{deudon_highres-net_2020} was the first approach to tackle the different problems in MISR (input co-registration, fusion, and registration-at-the-loss) in an end-to-end manner, and with a small memory footprint (due to its reused fusion operator in the low-res domain). Since then, several deep learning approaches with refined architectures have repeatedly beaten the state-of-the-art in the PROBA-V "post-mortem`` leaderboard; most notably \cite{salvetti2020multi}.

\subsubsection{HighRes-net} \label{subsubsec:highresnet}

This work applies the modified HighRes-net architecture of \cite{deudon_highres-net_2020} to the S-2 and PlanetScope RGB images described in sections \ref{subsec:sentinel2} and \ref{subsec:spacenet7}.
For a complete description of HighRes-net we refer the reader to the original paper \cite{deudon_highres-net_2020}.
Table \ref{tab:highresnet} is a modular breakdown of the HighRes-net architecture. The hallmark feature of HighRes-net is its shared fusion operator that can be pairwise-applied recursively on an arbitrary set of encodings of low-res revisits. This technique can be easily parallelized on a GPU, through careful use of the \texttt{torch.Tensor.view} operator, that treats the different encoding-pairs as instances of a mini-batch.

The next step, ShiftNet\footnote{A reduction of HomographyNet from 8 parameters for homographies to 2 for translations, see \cite{detone2016deep}.}, is not intrinsic to HighRes-net but an ancillary learned registration operator to account for the inevitable misalignments that are explained by shifts.

\subsection{Single-Image Super-Resolution}

Single-image super-resolution (SISR) has progressed significantly with the advent of deep learning, and has progressed even further with newer developments.
Initial approaches to super-resolution with deep architectures were CNN-based. First, SRCNN \cite{dong_image_2015} followed by FSRCNN \cite{dong_accelerating_2016}, which improved the speed of SRCNN along with a minor gain in performance. These approaches used some form of upsampling (transposed convolutions or bicubic upsampling) interspersed with convolutional layers to achieve the improved super-resolved imagery. The objective function used to train these networks is the mean square error between the high-resolution ground truth image and the super-resolved output image.

The work of~\cite{ledig_photo-realistic_2017} provided a significant step forward in terms of photo-realistic and perceptually pleasing super-resolution. They introduced a far better CNN based super-resolution network called Super-Resolution Residual Networks (SRResNet). SRResNet was deeper than prior works and included residual blocks. Additionally, instead of transposed convolutions or bicubic upsampling SRResNet made use of pixel shuffling to upsample the imagery. However, the major contribution of \cite{ledig_photo-realistic_2017}, was the introduction of a generative adversarial network for super-resolution (SRGAN) that achieved perceptually pleasing results. SRGAN uses SRResNet as the generator and a simple CNN for the discriminator. In  \cite{ledig_photo-realistic_2017}, they note that while SRResNet, trained with an MSE, achieves superior performance in terms of PSNR and SSIM to all other methods including SRGAN, SRGAN is able to achieve more perceptually pleasing results (capturing high-frequency content) as evaluated by Mean Opinion Score (MOS) of a panel of human evaluators.

More recently, there has been further work on improving super-resolution through improved GAN training \cite{wang2018esrgan, anwar_deep_2020, wang2020deep}, and flow-based approaches \cite{lugmayr2020srflow}.

\subsubsection{SRResNet}

While we examined many architectures as a SISR baseline to compare HighRes-Net against, we decided to restrict the comparison to SRResNet for a couple of reasons:
\begin{enumerate}
    \item It is amongst the best performing SISR methods in terms of PSNR and SSIM. 
    These are the primary metrics we will be evaluating our results on. We are interested in accuracy over perceptually pleasing results, and also had no time or budget to evaluate with panel of humans to obtain an MOS.
    \item SRResNet in terms of representational capacity is similar to that of HighRes-Net. Both networks rely upon residual blocks to encode the images. SRResNet does however make use of pixel shuffling instead of transpose convolution. Pixel shuffling has been shown to do better upsampling in \cite{ledig_photo-realistic_2017}.
\end{enumerate}

\begin{figure}
    \centering
    \includegraphics[width=0.72\textwidth]{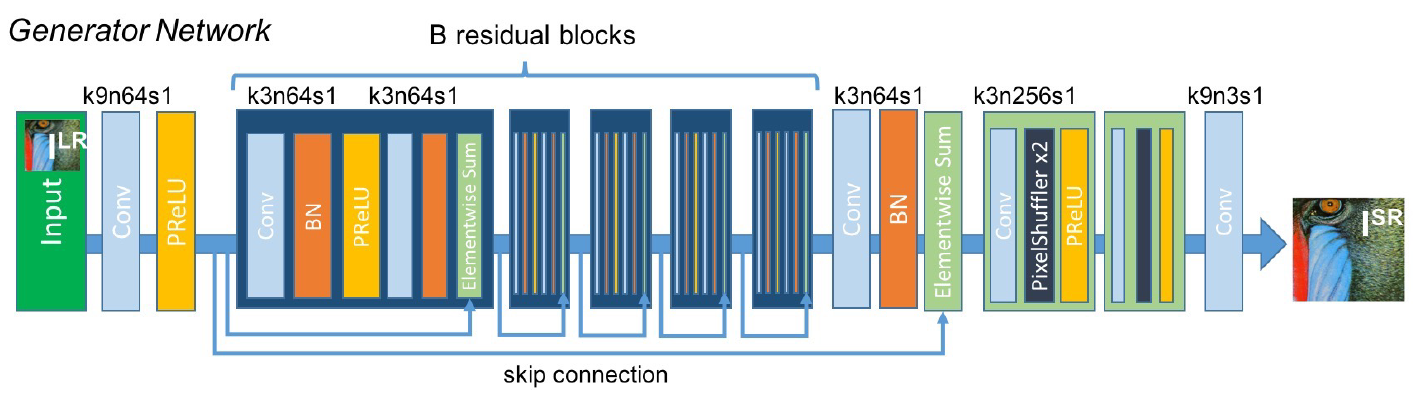}
    \caption{SRResNet model architecture as illustrated in \cite{ledig_photo-realistic_2017}.}
    \label{fig:SRResNet}
\end{figure}

\paragraph{Single-Image Selection}
Given the high Sentinel-2 revisit, one has to choose a revisit on which to do super-resolution using an SISR model. One could choose a random revisit, however, we found that choosing the best revisits indicated by the lack of clouds achieved optimal results.

\paragraph{Objective Function}
We evaluated three objective functions during training: Mean Square Error (MSE), Mean Absolute Error (MAE) and Structural Similarity Index Measure (SSIM). While MSE was able to obtain superior performance in terms of PSNR the imagery was far less sharp. Using SSIM as the objective function resulted in excellent PSNR performance, superior SSIM performance and imagery that was sharper and perceptually pleasing.

SRResNet's model architecture is illustrated in Figure \ref{fig:SRResNet}, the rest of the hyper-parameters used to train the network are described in the appendix~\ref{ap:hypersrrresnet}.

\section{Utility of MISR for Building Delineation} \label{sec:utilitymisr}

The interplay between super-resolution techniques and a range of downstream task performance (object detection, instance segmentation, semantic segmentation) remains largely unexplored, particularly in the context of satellite imagery. In \cite{shermeyer_effects_2019}, authors performed single image super-resolution using synthetically downscaled VHR WorldView-3 images. Afterwards, they compare the performance on an object detection task using images at different resolutions (real, downsampled and super-resolved images). They showed an increase in performance when using super-resolved images instead of artificially downsampled ones. This was far from a real-world scenario given the synthetic data used and moreover it was only assessed on a single task and dataset.

Detecting objects such as buildings from Sentinel-2 or even PlanetScope imagery, (resolution of 10m and 4m respectively in the visible spectrum), is a difficult task as these objects often cover a small amount of area in terms of pixels. In the case of buildings in urban areas, they are also densely packed making the task of delineating between buildings even more difficult. Increasing the spatial resolution in thus lead to better detection and delineation of these objects. 

In this study, we apply the previously trained super-resolution models to the downstream task of the SpaceNet dataset: semantic and instance segmentation of buildings.  We compare the models performance on the native (ground-truth) imagery, and super-resolved imagery of the same ground sampling distance and the low-resolution imagery with bicubic upsampling.

Ours is the first study to assess the utility of MISR, and the first study to assess the utility of SISR or MISR on real-world data, as opposed to synthetically downsampled imagery. 

\subsection{Downstream models and processing pipeline}


The state-of-the-art for building footprint segmentation has converged to a common algorithm: identify instances and extract polygons. The identify instance step has been approached as instance segmentation task, in which instances are directly obtained from an instance segmentation network, or more commonly as a semantic segmentation task, in which instances are obtained using the connected components of the mask predicted by a semantic segmentation network. In both approaches, a polygon is then extracted for each instance by a simple vectorization routine.

We follow the second avenue (semantic segmentation), as this was the approach used among all four winners of the SpaceNet-7 challenge.

\subsubsection{Input imagery}\label{subsec:inputconf}

As input imagery we use the MISR and SISR models outputs, trained as previously described in section \ref{subsec:MISR_for_EO}, in addition to a number of baselines.

\begin{enumerate}
    \item Bicubic upsampling: The clearest and less cloudy revisit of Sentinel-2 imagery was bicubicly upsampled.
    \item SISR: The clearest revisit of low-resolution Sentinel-2 imagery was fed through a trained SRResNet.
    \item Multi-Image Upsampling: The clearest four (4) revisit of low-resolution Sentinel-2 imagery was concatenated and bicubicly upsampled.
    \item MISR: All revisits of the low-resolution Sentinel-2 imagery were fed through HighRes-Net.
    \item PlanetScope: The high-resolution PlanetScope imagery.
\end{enumerate}

\subsubsection{Semantic Segmentation}

Semantic segmentation is the task of assigning each pixel a semantically meaningful label. In this setting, it is specifically the task of assigning whether a pixel belongs to a building or not. In this work, the aim of this evaluation is to compare the same semantic segmentation model using the different inputs referred before to assess the utility of the super-resolved images. For this purpose, we chose to use HRNet~\cite{hrnet} as the semantic segmentation model which is trained using all the input configurations of subsec.~\ref{subsec:inputconf}. We chose HRNet because it is the network used by the winning solution in the SpaceNet-7 challenge, and is amongst the state of the art networks as evaluated on variety of benchmarks~\cite{hrnet_v2}. The architecture for the network is illustrated in figure~\ref{fig:hrnet}, we list in appendix~\ref{ap:hyperhr} the hyper-parameters that we used for training the models; note that we did not perform a comprehensive hyper-parameter sweep.


\begin{figure}
    \centering
    \includegraphics[width=0.72\textwidth]{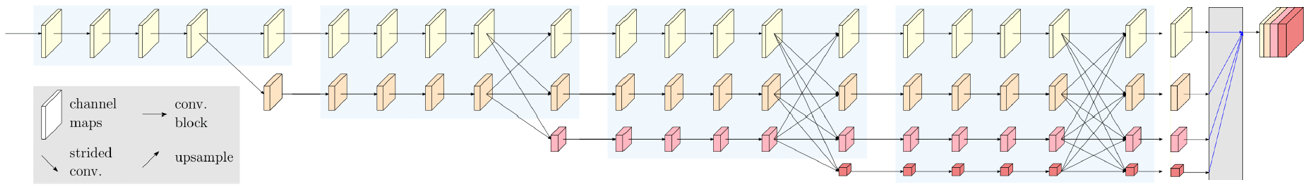}
    \caption{HRNet Architecture: high-resolution representations for Semantic Segmentation \cite{hrnet}.}
    \label{fig:hrnet}
\end{figure}

\paragraph{Semantic segmentation metrics}
The primary metric of semantic segmentation is Intersection-over-Union (IoU), which is evaluated for every class. The mean IoU over all the classes is the final reported metric.  IoU is measure of overlap between the predicted per-pixel segmentation and the ground truth divided by the area of union between this segmentation and the ground truth. IoU can be calculated as follows using either set or confusion matrix notation:
\begin{equation}
    IoU = \frac{A \cap B}{A \cup B} = \frac{TP}{TP+FP+FN}
\end{equation}
where $A$ is the predicted mask of the buildings and $B$ is the ground-truth mask.

\subsubsection{Instance Segmentation}

In the instance segmentation task, we aim to delineate each building instance. We do this through a simple process of vectorization that derives polygons using the connected components of the output of the semantic segmentation model. This is the standard method of polygon extraction and has been used within the SpaceNet challenge along with industrial uses by the likes of Microsoft for their maps.

\paragraph{Instance segmentation metrics}

The primary metrics for instance segmentation are precision, recall and F1. These metrics are calculated on the produced vector instances (polygons) rather than in the pixels. For a positive detection (true positive), the IoU between the prediction and ground-truth polygons must be greater than a threshold.
For this particular dataset, the IoU threshold in the competition was set at 0.25. Consequently, we used that threshold in our evaluation too.

\section{Color matching HighRes-net for Sentinel-2 MISR} \label{subsec:spectralconsistency}

The Sentinel-2 level 2A products consist of ortho-corrected images of BOA reflectance (see sec.~\ref{subsec:sentinel2}). There are several land and ocean S2 applications that rely on well-calibrated reflectance to provide meaningful outputs (see for instance~\cite{ruescas_machine_2018,wolanin_estimating_2019,svendsen_joint_2018}). However, the proposed MISR training scheme it is optimized to minimize disagreement between the PlanetScope and the super-resolved output image. Since PlanetScope images have a different color calibration than S2 (PlanetScope images from the SpaceNet-7 dataset are not BOA corrected), we experienced that the SR images {\it look like} PlanetScope images in terms of color (see Figure~\ref{fig:s2planet} to appreciate the difference in color between S2 and PlanetScope). Hence, by using the proposed MISR model, we loose the radiometric calibration of S2. In this section we propose a modification of the MISR model and training procedure to produce spectrally-consistent S2 super-resolved images. 

\begin{figure}
    \centering
    \includegraphics[width=.72\textwidth]{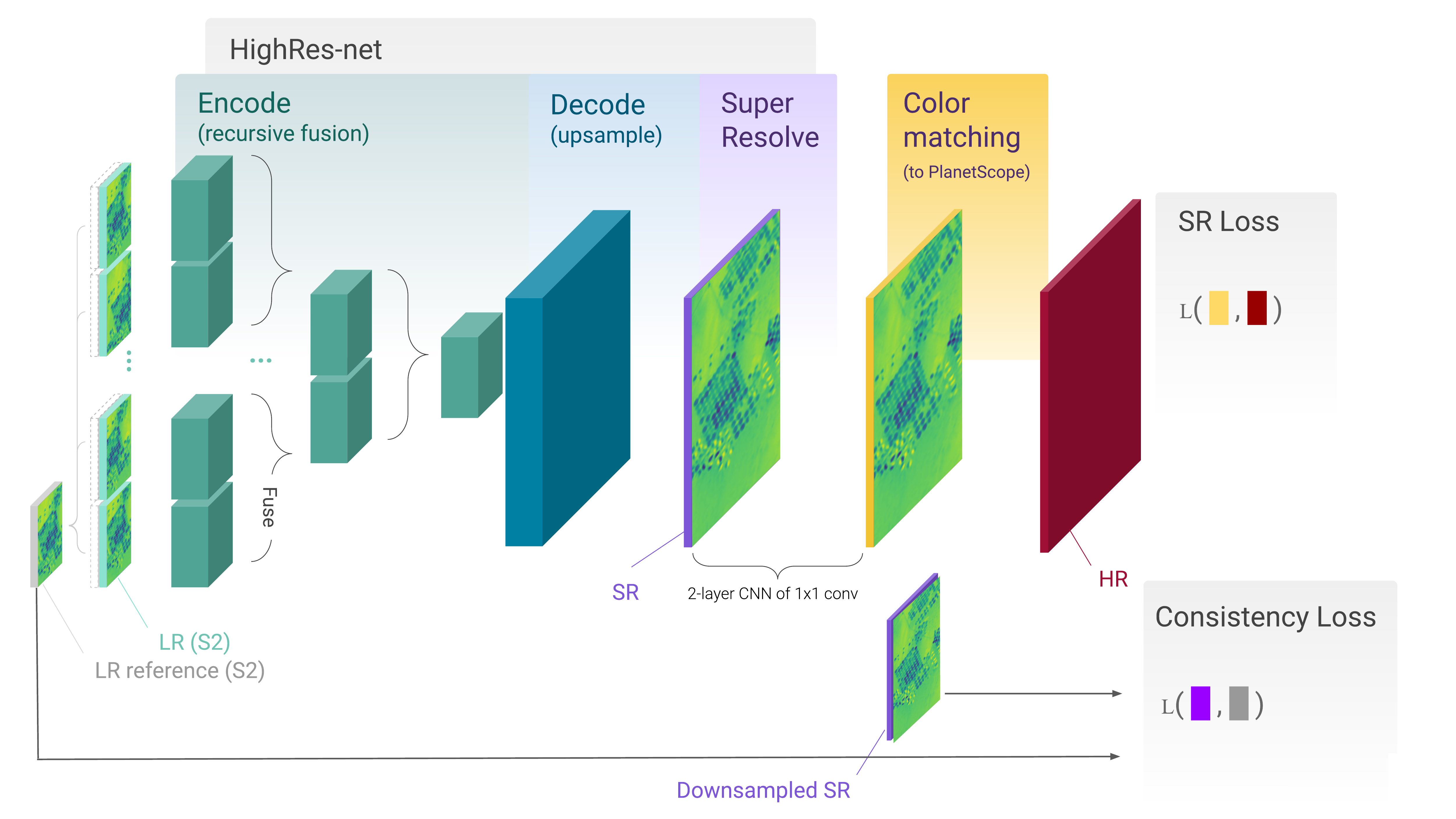}
    \caption{HighRes-net with a consistency loss between a downscaled version of the SR output and the low-res reference frame.}
    \label{fig:misrconsistency}
\end{figure}

Figure~\ref{fig:misrconsistency} shows a diagram of the modified forward pass with the consistency loss. On the top of the image we have HighRes-Net as proposed in \cite{deudon_highres-net_2020-1}: this network fuses a set of low-res revisits using the encoder (recursive fusion) and afterwards it applies the decoder to the fused feature-maps (upsample). However this output (SR in the figure), instead of being compared with the high-resolution image to compute the loss, it is further processed in two losses that are averaged together. For the first loss, that we called (color/spectral) \emph{consistency loss}, we downsample the SR image back to the low-res size and compare it with the S2 reference frame. For the second loss, called \emph{super-resolution loss}, we apply a pixel-by-pixel fully connected neural network (implemented as 2 layers of 1x1 convolutions) to produce the {\it PlanetScope like} output. We refer to this network as \emph{color matching network}. By using this additional network we seek to disentangle MISR (controlled by HighRes-Net) from color matching (learn by the color matching network). We were inspired by the works of \cite{tasar_colormapgan_2020} and \cite{mateo-garcia_cross-sensor_2020} that use similar ideas in the context of domain adaptation.

\subsection{Implementation Details}

For the consistency loss, we used the MSE between the downsampled image and the low-res reference, the downsampling step uses adaptive average pooling. The color matching network consists of a $1 \times 1$ convolutional layer with 64 output channels, a ReLU activation and a $1 \times 1$ conv layer back to 3 output channels. This output is shifted using the ShiftNet network, explained in sec. \ref{subsubsec:highresnet}, and compared with the PlanetScope high-res image using an SSIM loss. Finally, the consistency loss and the super-resolution loss are averaged together using a convex combination of $0.9$ and $0.1$ respectively.


\section{Results}
\subsection{Multi-Image Super-Resolution Results} \label{sec:results}

\paragraph{Quantitative results}
This section focuses on performance of the models in terms of Peak Signal to Noise Ratio (PSNR) and Structural Similarity Index Measure (SSIM). These are the most commonly used metrics for measuring super-resolution performance and image reconstruction quality.

Table \ref{tab:sr_perf} shows the super-resolution performance of our evaluated models with the \textit{within-scene} split described in \ref{subsubsec:splits}. 
\textit{Bicubic} method indicates the scores if LR images are just upscaled using bicubic interpolation to match the HR image size.

\begin{table}[!ht]
    \caption{
        \small Average PSNR  and SSIM scores. Higher scores are better.
    }
    \label{tab:sr_perf}
    \centering
    \begin{tabular}{lllllllll}
    \toprule
    Subset &  \multicolumn{2}{c}{Bicubic} & \multicolumn{2}{c}{SRResNet} & \multicolumn{2}{c}{HighResNet} \\
    {} & PSNR & SSIM & PSNR & SSIM & PSNR & SSIM \\
    \midrule
    Train & 17.85 & 0.612 & 27.90 & 0.827 & \textbf{30.16} & \textbf{0.88}\\
    Validation & 18.11 & 0.70 & 26.06 & 0.78 & \textbf{28.52} & \textbf{0.84}\\
    Test & 18.54 & 0.70 & 26.83 & 0.80 & \textbf{29.40} & \textbf{0.85}\\
    \bottomrule
    \end{tabular}
\end{table}

In addition, we tested the same model applied to acquisitions in a different time period but over the same regions. In Table \ref{tab:sr_perf_time}, we notice a drop in performance across all methods, but the super-resolution models still exceed the performance of the bi-cubic upsampling method. Additionally, we still see that MISR outperforms SISR.

\begin{table}[!ht]
    \caption{\small Average PSNR and SSIM scores for testing on a different time period.}
    \label{tab:sr_perf_time}
    \centering
    \begin{tabular}{lllllllll}
    \toprule
    Subset &  \multicolumn{2}{c}{Bicubic} &    \multicolumn{2}{c}{SRResNet} & \multicolumn{2}{c}{HighResNet} \\
     {} & PSNR & SSIM & PSNR & SSIM & PSNR & SSIM \\
    \midrule
    Train & 17.27 & 0.63 & 22.66 & 0.74 & \textbf{23.28} & \textbf{0.76}\\
    Validation & 18.14 & 0.69 & 22.49 & \textbf{0.73} & \textbf{22.56} & \textbf{0.73}\\
    Test & 16.99 & 0.66 & 22.12 & 0.73 & \textbf{23.10} & \textbf{0.77}\\
    \bottomrule
    \end{tabular}
\end{table}

\subsection{Building Detection Results}

In this subsection, we present the results of the downstream tasks. Due to the manner in which the tasks are completed, they are very much linked.

\subsubsection{Semantic Segmentation}

In the semantic segmentation results, we see that, as expected, the model using MISR exceeds the performance of the bicubic upsampling and SISR. However, the model using multiple revisits (with bicubic upsampling) is able to match the performance of the MISR model. This result shows the utility of using multiple revisits, while at the same time showing that MISR (or at least our particular implementation) does not necessarily result the optimal representation for performance on the downstream task. Optimizing directly using the best 4 revisits results in equal or better performance.

Additionally, we see that SISR underperforms against bicubic upsampling. We suspect this is likely due to SISR introducing artifacts which results in the lower performance. 

Finally, as expected, the PlanetScope imagery ground-truth resulted in a model with far superior performance than any of the other methods.

\begin{table}[!ht]
    \caption{Performance of semantic segmentation models trained with different input imagery. On-par top accuracy shown in bold.}
    \label{tab:semanticresults}
    \centering
    \begin{tabular}{lrr}
    \toprule
    Model                                               & IoU (val) & IoU (test) \\
    \midrule
    S-2 (best revisit): Bicubic                  & 0.60     & 0.60      \\ 
    S-2 (best revisit): SISR                     & 0.58     & 0.56      \\ 
    S-2 (best 4 revisits): concat + bicubic      & \textbf{0.62}     & \textbf{0.62}      \\ 
    S-2: MISR (all revisits)                     & \textbf{0.61}     & \textbf{0.60}      \\
    \midrule
    PlanetScope                                  & 0.66     & 0.69     \\
    \bottomrule
    \end{tabular}
\end{table}

\begin{figure}[!ht]
    \centering
    \includegraphics[width=0.72\textwidth]{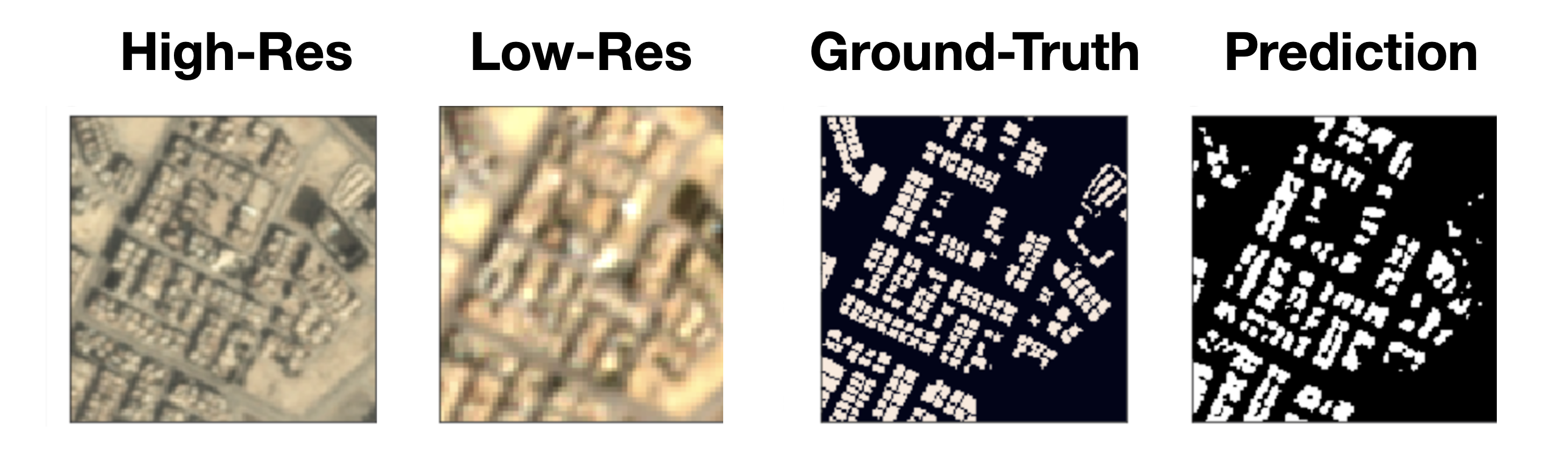}
    \caption{A sample result of semantic segmentation on SpaceNet 7 dataset using MISR imagery from Sentinel-2.}
    \label{fig:semanticex}
\end{figure}

An example of the resulting semantic segmentation output given the inputs is shown in Figure \ref{fig:semanticex}.

\subsubsection{Instance Segmentation}

The results on the instance segmentation task largely mimic the results seen in the semantic segmentation results. We can, however, see that performance margin grows between the model using the PlanetScope imagery and those using Sentinel-2, indicating the performance of the fine point accuracy required for correct delineation for instance segmentation on this particularly dataset.

We also see that precision is generally lower with the super-resolution imagery than with their equivalent bicubicly upsampled baselines. This indicates the super-resolution models are producing predictions that are less pure.

One important note to make is that we used all revisits with the MISR model (including cloudy revisits), while the bicubicly upsampled model only used the best four revisits. While the MISR model was shown to be relatively robust to this, an avenue for further investigation should include using simply the best four revisits for the MISR model.

\begin{table}[!ht]
    \caption{Performance of instance segmentation algorithm with different input imagery on the test set}
    \label{tab:instancesegmentation}
    \centering
    \begin{tabular}{lrrr}
    \toprule
    Model (IoU Threshold: 0.25)                           & Precision & Recall & F1   \\
    \midrule
    S-2 (best revisit): Bicubic                    & 0.24      & 0.56   & 0.33 \\ 
    S-2 (best revisit): SISR                       & 0.21      & 0.56   & 0.30 \\ 
    S-2 (best 4 revisits): concat + bicubic        & \textbf{0.28}      & \textbf{0.57}   & \textbf{0.37} \\ 
    
    S-2: MISR (all revisits)                       & \textbf{0.26}      & \textbf{0.57}   & \textbf{0.36} \\
    PlanetScope                                    & 0.41      & 0.68   & 0.51 \\
    \bottomrule
    \end{tabular}
\end{table}

Overall, we saw that MISR was shown to be useful for the resulting downstream tasks of semantic and instance segmentation. These results suggest that MISR images encode the information contained in the Sentinel-2 time series. 

\subsection{Colour Consistency Results}
\begin{figure}[!ht]
    \centering
    \includegraphics[width=0.8\linewidth]{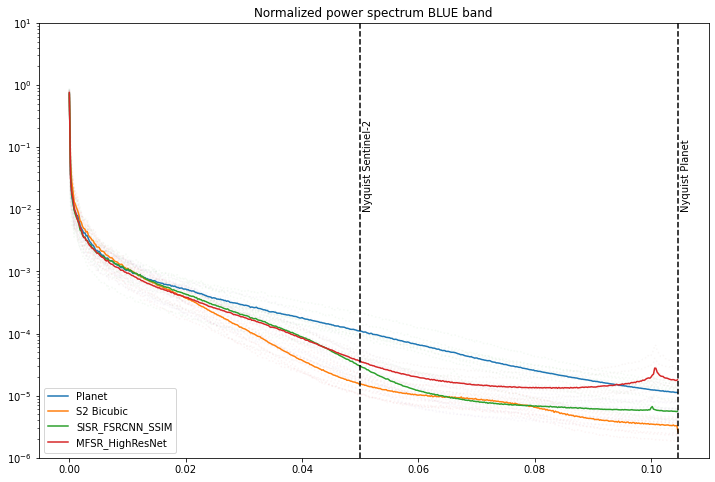}
    \caption{Averaged power spectrum of PlanetScope, Sentinel-2 and Sentinel-2 super-resolved images over the train and validation dataset. We can see that super-resolved images using MISR and SISR (green and red) have more high-frequency components than Sentinel-2 (orange). }
    \label{fig:spatialfourier}
\end{figure}

\textbf{Averaged spatial Fourier power spectrum}. 
Figure~\ref{fig:spatialfourier} shows the spatial Fourier power spectrum averaged over the images in the test and validation datasets. In particular, we show the spectrum of PlanetScope images, Sentinel-2 images upscaled with bicubic interpolation and the super-resolved images of the MISR and SISR methods. We can see that super-resolved images (red and green) exhibit a larger amount of higher-frequency components than Sentinel-2 images (orange). This shows that the SR models are adding higher-frequency content to the output. On the other hand, still, the amount of high-frequency information is lower than in PlanetScope (blue). 

\begin{figure}[ht]
    \centering
    \includegraphics[width=.8\linewidth]{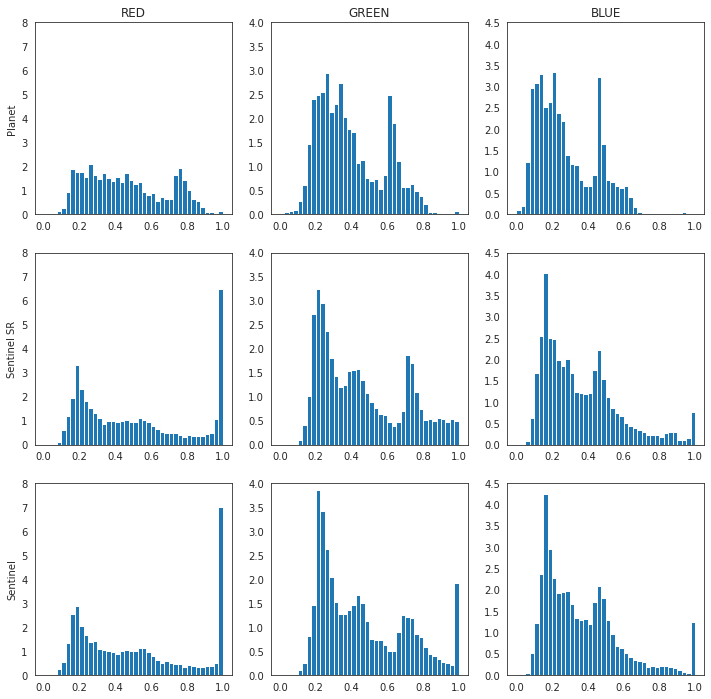}
    \caption{Normalized histograms of the image values over the test and val dataset for each of the RGB bands (columns). First row shows the values of PlanetScope images, second row shows the values of the super-resolved Sentinel-2 images using the consistency loss. Third row shows the values of the Sentinel-2 images. We see that, using the color consistency loss, the color distribution of Sentinel-2 is maintained in the super-resolved images.}
    \label{fig:consistency_histograms}
\end{figure}

\textbf{Radiometric consistency}. 
In order to evaluate the radiometric consistency method, Figure \ref{fig:consistency_histograms} shows the histograms over the test and val datasets of each of the RGB bands. In these histograms we can see that the color distribution of PlanetScope (first row) is very different of the colors of Sentinel-2 (last row). In the middle we show the histograms of the Sentinel-2 super-resolved images using the MISR method with consistency loss (sec.\ref{subsec:spectralconsistency}). We can see that the color distribution of the SR images matches the color distribution of Sentinel-2.

\subsection{Qualitative results}
\begin{figure}[ht]
    \centering
    \subfloat[Low-res (S-2, 10m)]{
        \includegraphics[width=0.31\linewidth, trim={0ex 0ex 0ex 0ex}, clip]{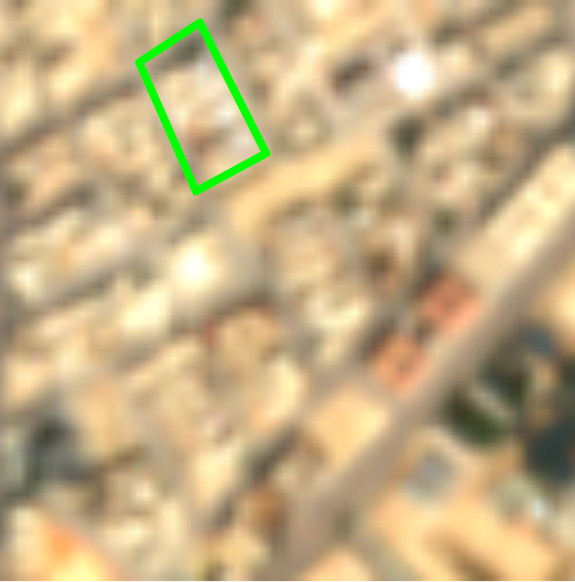}
    	\label{fig:MISR_example_lr}
    }
    \subfloat[Super-res (4.7m)]{
        \includegraphics[width=.31\linewidth, trim={0ex 0ex 0ex 0ex}, clip]{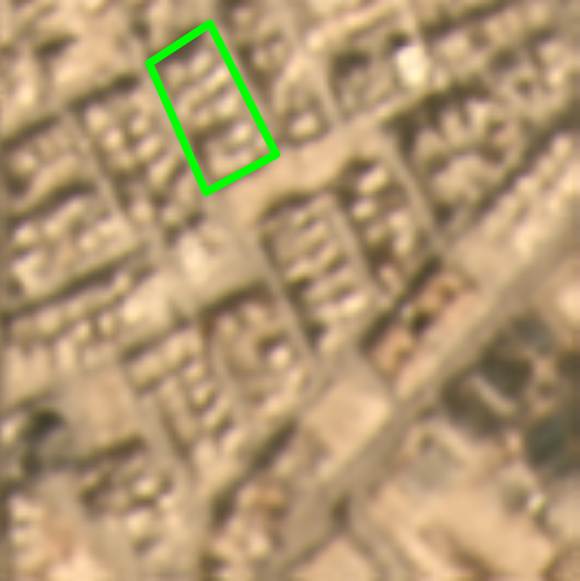}
    	\label{fig:MISR_example_sr}
    }
    \subfloat[High-res (PlanetScope, 4.7m)]{
        \includegraphics[width=.31\linewidth, trim={0ex 0ex 0ex 0ex}, clip]{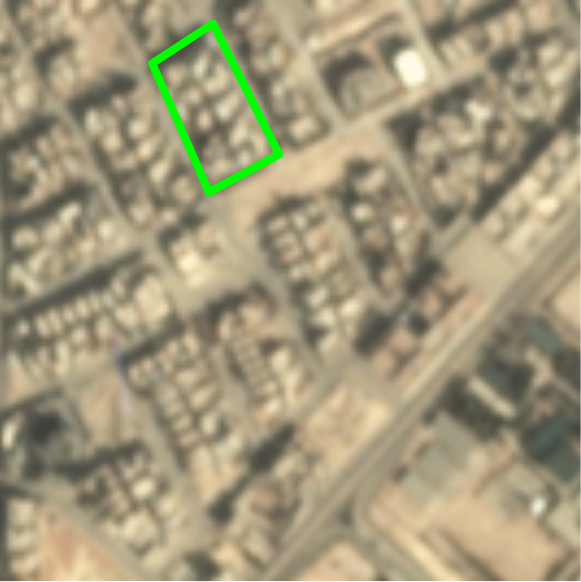}
    	\label{fig:MISR_example_hr}
    }
    \\
    \subfloat[]{
        \includegraphics[width=0.31\linewidth, trim={0ex 0ex 0ex 0ex}, clip]{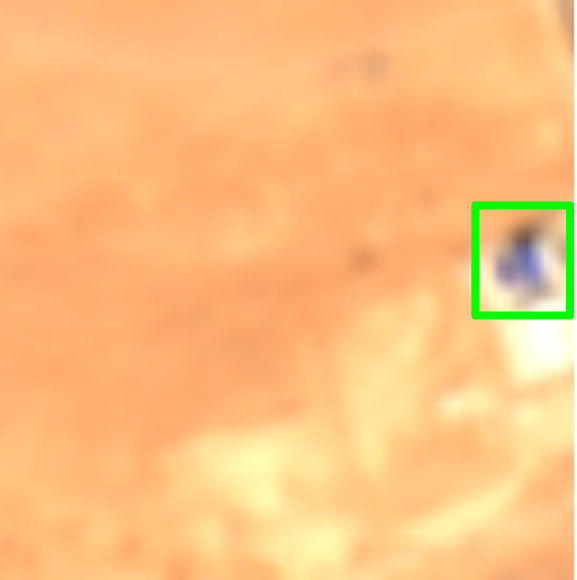}
    	\label{fig:MISR_example2_lr}
    }
    \subfloat[]{
        \includegraphics[width=.31\linewidth, trim={0ex 0ex 0ex 0ex}, clip]{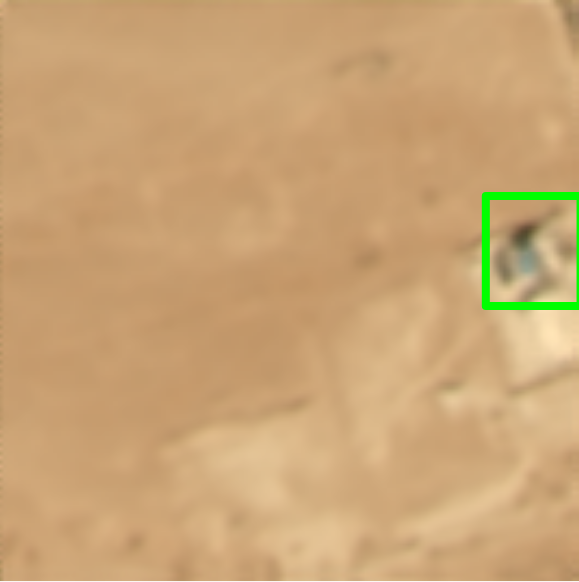}
    	\label{fig:MISR_example2_sr}
    }
    \subfloat[]{
        \includegraphics[width=.31\linewidth, trim={0ex 0ex 0ex 0ex}, clip]{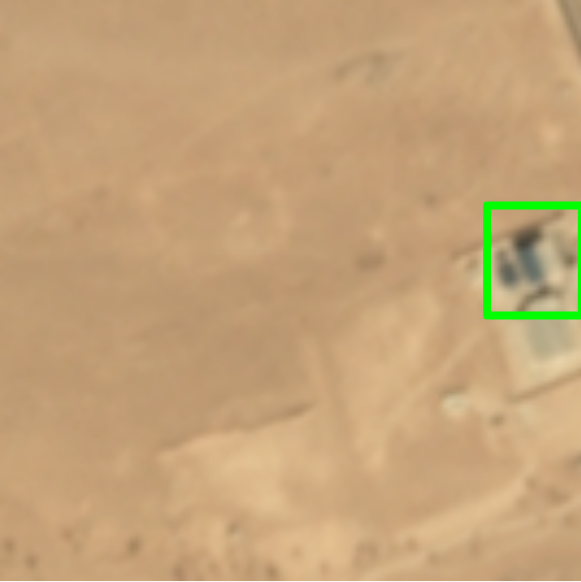}
    	\label{fig:MISR_example2_hr}
    }
    \caption{
        \small Patch from a validation-set scene. How many buildings lie within the green polygon? Being more than just a pretty picture, the super-resolved output of HighRes-net \ref{fig:MISR_example_sr} also better delineates the buildings in urban scenes, hence enabling downstream tasks like building segmentation, with improved accuracy compared to prediction on a single S-2 image. This is evidenced qualitatively by the fact that the manual count of buildings in the {\color{green}green} polygon in \ref{fig:MISR_example_sr} is easier to perform than in \ref{fig:MISR_example_lr}.
        \ref{fig:MISR_example_lr}---\ref{fig:MISR_example2_hr} Note that in both examples/rows, the spectra of the super-res output is similar to the high-res PlanetScope reference --- an undesirable side-effect if the spectral information of the source low-res instrument is better than that of the high-res instrument.
    }
    \label{fig:MISR_example}
\end{figure}

\begin{figure}[ht]
    \centering
    \subfloat[LR (S2, 10m)]{
        \includegraphics[width=0.23\linewidth, trim={0ex 0ex 0ex 0ex}, clip]{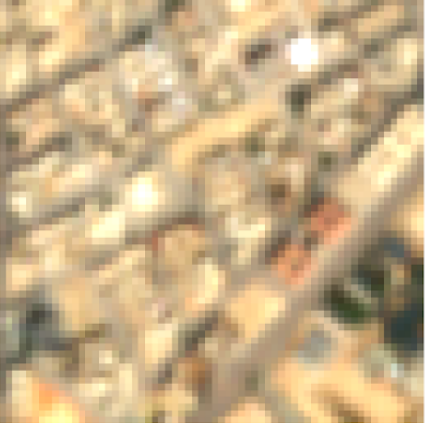}
    	\label{fig:consistency_example_lr}
    }
    \subfloat[SR with S2 spectra (4.7m)]{
        \includegraphics[width=.23\linewidth, trim={0ex 0ex 0ex 0ex}, clip]{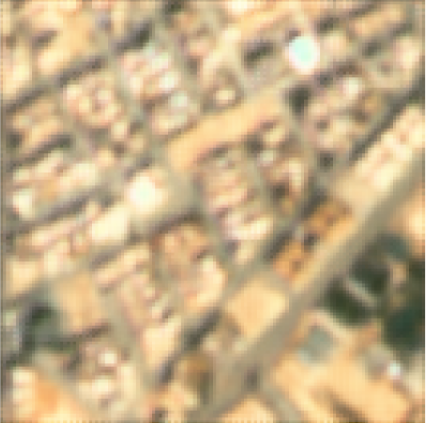}
    	\label{fig:consistency_example_srs2}
    }
    \subfloat[SR with PS spectra (4.7m)]{
        \includegraphics[width=.23\linewidth, trim={0ex 0ex 0ex 0ex}, clip]{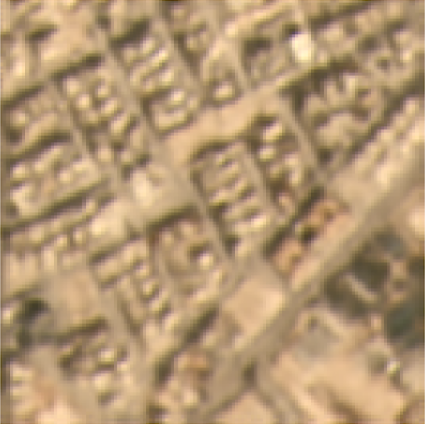}
    	\label{fig:consistency_example_srps}
    }
    \subfloat[HR ground-truth (4.7m)]{
        \includegraphics[width=.23\linewidth, trim={0ex 0ex 0ex 0ex}, clip]{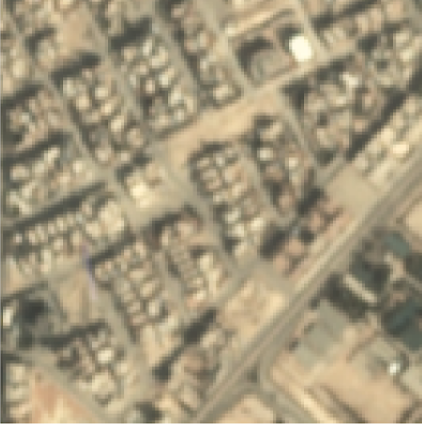}
    	\label{fig:consistency_example_hr}
    }
    \\
    \subfloat[]{
        \includegraphics[width=0.23\linewidth, trim={0ex 0ex 0ex 0ex}, clip]{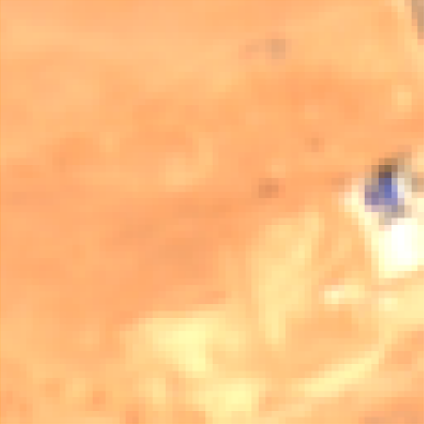}
    	\label{fig:consistency_example2_lr}
    }
    \subfloat[]{
        \includegraphics[width=.23\linewidth, trim={0ex 0ex 0ex 0ex}, clip]{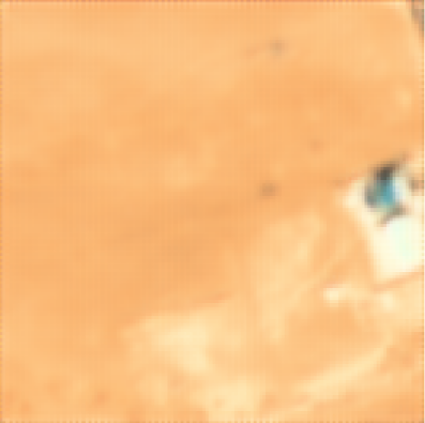}
    	\label{fig:consistency_example2_srs2}
    }
    \subfloat[]{
        \includegraphics[width=.23\linewidth, trim={0ex 0ex 0ex 0ex}, clip]{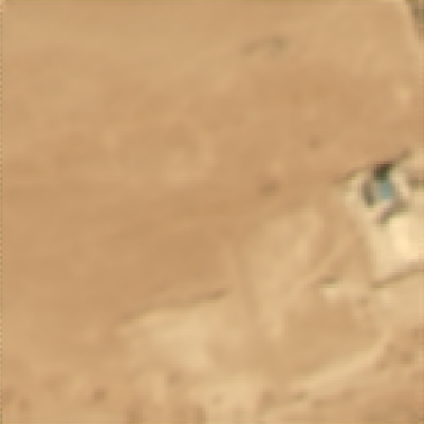}
    	\label{fig:consistency_example2_srps}
    }
    \subfloat[]{
        \includegraphics[width=.23\linewidth, trim={0ex 0ex 0ex 0ex}, clip]{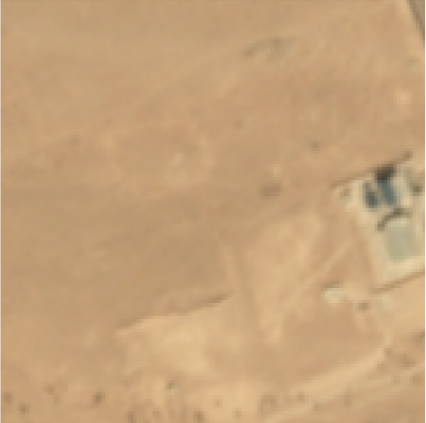}
    	\label{fig:consistency_example2_hr}
    }
    \\
    \subfloat[]{
        \includegraphics[width=0.23\linewidth, trim={0ex 0ex 0ex 0ex}, clip]{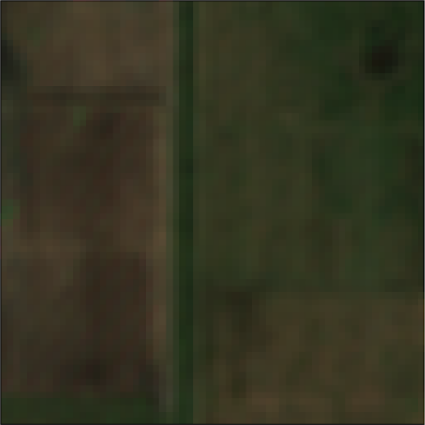}
    	\label{fig:consistency_example3_lr}
    }
    \subfloat[]{
        \includegraphics[width=.23\linewidth, trim={0ex 0ex 0ex 0ex}, clip]{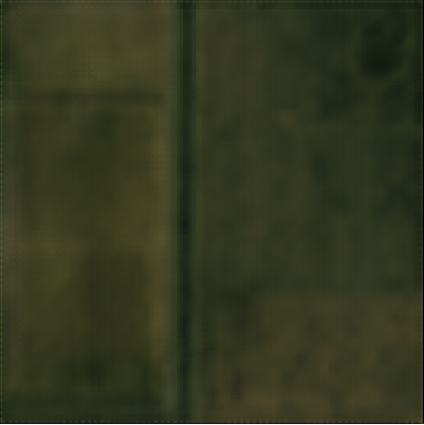}
    	\label{fig:consistency_example3_srs2}
    }
    \subfloat[]{
        \includegraphics[width=.23\linewidth, trim={0ex 0ex 0ex 0ex}, clip]{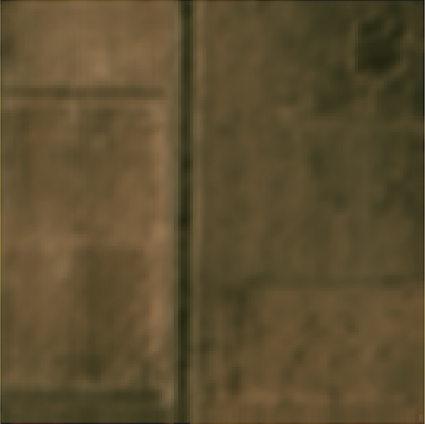}
    	\label{fig:consistency_example3_srps}
    }
    \subfloat[]{
        \includegraphics[width=.23\linewidth, trim={0ex 0ex 0ex 0ex}, clip]{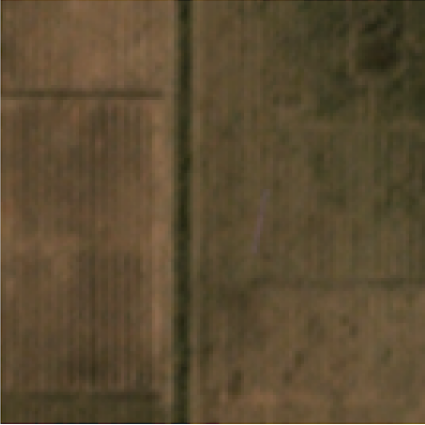}
    	\label{fig:consistency_example3_hr}
    }
    \caption{
        \small Example outputs of HighRes-net with the consistency loss, to preserve the spectra of Sentinel-2. 
        \ref{fig:consistency_example_lr}---\ref{fig:consistency_example_hr}: Building are best delineated in \ref{fig:consistency_example_srps} but the spectra are no longer consistent with Sentinel-2.
        \ref{fig:consistency_example_srs2} presents a trade-off that still delineates buildings better than \ref{fig:consistency_example_lr} while being consistent to S-2 spectra.
        \ref{fig:consistency_example2_lr}---\ref{fig:consistency_example2_hr}: Example patch from Figure \ref{fig:MISR_example}, now with S-2 spectra preserved in \ref{fig:consistency_example2_srs2}.
        \ref{fig:consistency_example3_lr}---\ref{fig:consistency_example3_hr}: Example of a failure in an agricultural scene, where HighRes-net with the consistency loss \ref{fig:consistency_example3_srs2} has failed to capture the lack of vegetation originally shown in the left side of \ref{fig:consistency_example3_lr}. We suspect this is because the high-res ground-truth contains vegetation. In other words, this instance highlights the compromise between the spectra of S-2 and content of PlanetScope.
    }
    \label{fig:consistency_example}
\end{figure}

Figure \ref{fig:MISR_example} shows a few examples of the visual quality of our MISR approach and Figure \ref{fig:consistency_example} shows the effect of the consistency loss in HighRes-net on a few test-set instances. Firstly, image~\ref{fig:MISR_example} shows a Sentinel-2 low res, a super resolved image using HighRes-net and the PlanetScope HR image. Particularly, the first row shows an urban area where the super-resolved image (middle) is significantly sharper than the low-res Sentinel-2 image (left); in this image, it is clear that counting buildings should be easier in the former than in the later. Figure \ref{fig:consistency_example} shows examples of SR with colour consistency (second column); we can see that the proposed model enhance the spatial resolution of Sentinel-2 while preserving the color content.  The output on the third column shows the output after the color matching step, which, as we see it is more similar to the PlanetScope ground truth.

\section{Discussion and conclusions}
In this work, we propose the first MISR model for multi-spectral imagery adapting the HighRes-net model of Deudon et al.~\cite{deudon_highres-net_2020}. We demonstrate this model over a new compiled dataset of high-res PlanetScope images (4.77m) and Sentinel-2 lower-res (10m) time series over 46 different locations using the SpaceNet-7 dataset as a base. Using this data, we show that the HighRes-net model produce better super-resolution images than SISR or bicubic upsampling in terms of the PSNR and SSIM metrics.  These results are aligned with the outputs of the PROBA-V contest organized by the ESA that highlighted the potential of deep learning based super-resolution for Earth observation. 

Additionally, we propose a modification of the HighRes-net architecture and training procedure to deal with the domain shift between PlanetScope and Sentinel-2 images. Taking into account the spectral differences between instruments has been overlooked by previous super-resolution works (e.g. in the SISR work of \cite{salgueiro_romero_super-resolution_2020} authors ignore this issue when working with WorldView and Sentinel-2 images). Nevertheless, the radiometric calibration of Sentinel-2 images is higher (12-bit depth) than the PlanetScope data (8-bit) and Sentinel-2 atmospheric correction is also more accurate due to the dedicated atmospheric correction bands (B1, B9 and B10). Hence, surface reflectance BOA values for Sentinel-2 images are more accurate than in PlanetScope. With our proposed solution we seek to have the best of both worlds: high spatial resolution and good radiometrically calibrated data. 

Another important contribution of this work is the demonstration of MISR in a downstream task. In particular, we showed that MISR-fused revisits produce training images that yield better performance of building segmentation models, compared to SISR and bicubic upsampling, in terms of IoU, Recall / Precision / F1 scores. Interestingly, SISR performs worse than bicubic upsampling, possibly due to artefacts caused by the deep SISR image generator. We also tried a more simplified approach of fusion, by concatenating multiple revisits on the channels dimension, and forward-passing them with a Fully Convolutional Net\footnote{A CNN with no pooling on the spatial dimensions.} for segmentation. We found that even with this rather simplistic approach, multiple revisits still assist the segmentation task (although we are using the 4 best revisits while HighRes-net receives all the inputs and have to learn to discard the cloudy acquisitions).

\subsection{Future work} 
To attempt a fully multi-spectral MISR approach, on all bands of Sentinel-2, ideally a higher resolution reference with the same bands would be needed, at least in a supervised learning setting. Further research into unsupervised MISR is needed to unlock the super-resolution potential of any revisit archive, without depending on near co-located and co-temporal higher resolution imagery.

Although the colour consistency loss shows promising results in the preservation of Sentinel-2 spectra, we conclude that a more flexible approach is needed for fusing revisits from dynamic scenes, by attending and fusing features that are static with respect to a certain \textit{anchor} revisit, possibly chosen by the user. This anchor revisit can be any one within the set of available revisits, and the \textit{anchored fusion} model should enhance only the parts of the image that are static with respect to the anchor image, and ignore any dynamic features (e.g. due to weather, vegetation, urban development).


\end{paracol}
\vspace{6pt} 



\authorcontributions{Conceptualization, M.R., F.K. and Y.G.; methodology, M.R., F.K. G.M.G, Y.G. ; data curation, M.R., F.K. G.M.G; writing---original draft preparation, M.R., F.K.; writing---review and editing, M.R., F.K. G.M.G, L.G.C. Y.G.;  supervision, F.K, Y.G. L.G.C..
}

\funding{This work has been enabled by \href{https://fdleurope.org}{Frontier Development Lab (FDL) Europe}, a public partnership between the European Space Agency (ESA) at Phi-Lab (ESRIN), Trillium Technologies and the University of Oxford; the project has been also supported by Google Cloud. G.M.-G. and L.G.-C. are funded by the Spanish ministry of Science DOI MCIN/AEI/10.13039/501100011033/ project TEC2016-77741-R.}




\dataavailability{The PlanetScope images used in this work are publicly available in \url{https://spacenet.ai/sn7-challenge/}. Sentinel-2 images were obtained from the Copernicus Open Access Hub \url{https://scihub.copernicus.eu/}. The co-registered dataset produced in this study will be made publicly available in \url{http://spaceml.org/}.} 

\acknowledgments{This work has been enabled by \href{https://fdleurope.org}{Frontier Development Lab (FDL) Europe}, a public partnership between the European Space Agency (ESA) at Phi-Lab (ESRIN), Trillium Technologies and the University of Oxford; the project has been also supported by Google Cloud.
The authors would like to thank the support of James Parr and Jodie Hughes from the Trillium team and to Nicolas Longépé from ESA PhiLab for discussions and comments throughout the development of this work.}

\conflictsofinterest{The authors declare no conflict of interest.} 




\newpage
\appendixtitles{no} 
\appendixstart
\appendix
\section{Dataset Area of Interest Breakdown}
\begin{table}[!h]
    \caption{
        The subset (\red{45}) of SpaceNet-7 AOIs that we used in this work, acquired between December 2019 and January 2020. The high-level breakdown of the types of terrain contained in each scene shows the overall geo-diversity of the dataset.
        Left to right: \textit{\% clouds} is the average cloud coverage (SCL=9) across all revisits; \textit{desert, agri(culture), urban, veg(etation), bare (soils)} indicate the type of terrain; a \textit{usable} revisit is at least \%50 cloud-free; \textit{val/test} indicate whether the scene is part of the validation or testing dataset (\textit{NA}=\textit{not used}).
    }
    \label{tab:AOI}
    \centering
    \begin{tabular}{@{}clccccccclccc@{}}
    id & Scene & \%clouds & desert & agri & urban & veg & bare & \multicolumn{3}{c}{revisits} & val & test \\
     & &  &  &  &  &  &  & usable & total & \% &  & \\
    \midrule
    1 & 0358E-1220N\_1433\_3310 & 70 &  & 1 & 1 & 1 &  & 8 & 13 & 62 &  & 1 \\
    2 & 1389E-1284N\_5557\_3054 & 69 & 1 &  & 1 &  & 1 & 8 & 13 & 62 &  & 1 \\
    3 & 0361E-1300N\_1446\_2989 & 66 &  & 1 &  &  & 1 & 8 & 13 & 62 & 1 &  \\
    4 & 1848E-0793N\_7394\_5018 & 66 &  & 1 & 1 &  & 1 & 7 & 13 & 54 &  &  \\
    5 & 0357E-1223N\_1429\_3296 & 65 &  & 1 & 1 & 1 &  & 6 & 13 & 46 &  &  \\
    6 & 1716E-1211N\_6864\_3345 & 62 &  &  & 1 &  &  & 5 & 13 & 38 &  &  \\
    7 & 1025E-1366N\_4102\_2726 & 55 &  & 1 & 1 & 1 &  & 5 & 13 & 38 &  &  \\
    8 & 1672E-1207N\_6691\_3363 & 53 &  &  & 1 &  & 1 & 5 & 13 & 38 &  &  \\
    9 & 1298E-1322N\_5193\_2903 & 56 &  & 1 & 1 &  & 1 & 4 & 13 & 31 &  &  \\
    10 & 1014E-1375N\_4056\_2688 & 49 &  & 1 & 1 & 1 &  & 4 & 13 & 31 &  &  \\
    11 & 1703E-1219N\_6813\_3313 & 58 &  & 1 & 1 &  & 1 & 3 & 13 & 23 &  &  \\
    12 & 1617E-1207N\_6468\_3360 & 24 &  &  & 1 & 1 &  & 3 & 13 & 23 &  &  \\
    13 & 1439E-1134N\_5759\_3655 & 61 &  & 1 & 1 & 1 &  & 5 & 10 & 50 &  &  \\
    14 & 0566E-1185N\_2265\_3451 & 96 &  & 1 &  & 1 &  & 6 & 7 & 86 & 1 &  \\
    15 & 0586E-1127N\_2345\_3680 & 83 &  & 1 & 1 & 1 &  & 6 & 7 & 86 &  & 1 \\
    16 & 1481E-1119N\_5927\_3715 & 81 &  & 1 & 1 & 1 &  & 6 & 7 & 86 &  &  \\
    17 & 0571E-1075N\_2287\_3888 & 81 &  &  & 1 & 1 &  & 6 & 7 & 86 &  &  \\
    18 & 1200E-0847N\_4802\_4803 & 80 &  & 1 & 1 & 1 &  & 6 & 7 & 86 &  &  \\
    19 & 1210E-1025N\_4840\_4088 & 95 &  & 1 & 1 & 1 &  & 5 & 7 & 71 &  &  \\
    20 & 1335E-1166N\_5342\_3524 & 84 & 1 &  & 1 &  & 1 & 5 & 7 & 71 &  &  \\
    21 & 1204E-1204N\_4819\_3372 & 74 & 1 &  & 1 &  & 1 & 5 & 7 & 71 &  &  \\
    22 & 0632E-0892N\_2528\_4620 & 67 & 1 &  & 1 &  & 1 & 5 & 7 & 71 &  &  \\
    23 & 1479E-1101N\_5916\_3785 & 67 &  & 1 &  & 1 &  & 5 & 7 & 71 &  &  \\
    24 & 0434E-1218N\_1736\_3318 & 84 & 1 &  &  &  &  & 4 & 7 & 57 &  &  \\
    25 & 1138E-1216N\_4553\_3325 & 71 &  &  & 1 &  & 1 & 4 & 7 & 57 &  &  \\
    26 & 0331E-1257N\_1327\_3160 & 68 &  & 1 &  & 1 &  & 4 & 7 & 57 &  &  \\
    27 & 1049E-1370N\_4196\_2710 & 59 &  &  & 1 & 1 &  & 4 & 7 & 57 &  &  \\
    28 & 1185E-0935N\_4742\_4450 & 33 &  & 1 & 1 & 1 &  & 2 & 7 & 29 &  &  \\
    29 & 0614E-0946N\_2459\_4406 & 29 &  &  & 1 & 1 &  & 2 & 7 & 29 &  &  \\
    30 & 0595E-1278N\_2383\_3079 & 34 & 1 &  &  &  &  & 1 & 7 & 14 & NA & NA \\
    31 & 1209E-1113N\_4838\_3737 & 100 & 1 & 1 & 1 &  & 1 & 6 & 6 & 100 & 1 &  \\
    32 & 0977E-1187N\_3911\_3441 & 100 & 1 &  & 1 &  & 1 & 6 & 6 & 100 &  &  \\
    33 & 1289E-1169N\_5156\_3514 & 99 &  &  & 1 &  & 1 & 6 & 6 & 100 &  &  \\
    34 & 0368E-1245N\_1474\_3210 & 67 &  &  & 1 &  & 1 & 6 & 6 & 100 &  &  \\
    35 & 1015E-1062N\_4061\_3941 & 100 &  &  & 1 & 1 &  & 5 & 6 & 83 &  &  \\
    36 & 1438E-1134N\_5753\_3655 & 92 &  &  & 1 & 1 &  & 5 & 6 & 83 &  & 1 \\
    37 & 1276E-1107N\_5105\_3761 & 91 &  & 1 & 1 &  & 1 & 5 & 6 & 83 & 1 &  \\
    38 & 1296E-1198N\_5184\_3399 & 87 & 1 &  & 1 &  & 1 & 5 & 6 & 83 &  &  \\
    39 & 0924E-1108N\_3699\_3757 & 67 &  &  & 1 &  &  & 4 & 6 & 67 &  &  \\
    40 & 0487E-1246N\_1950\_3207 & 98 &  & 1 &  & 1 &  & 3 & 6 & 50 &  &  \\
    41 & 1538E-1163N\_6154\_3539 & 63 &  &  & 1 & 1 &  & 3 & 6 & 50 &  &  \\
    42 & 1748E-1247N\_6993\_3202 & 57 &  &  & 1 &  & 1 & 3 & 6 & 50 &  &  \\
    43 & 1172E-1306N\_4688\_2967 & 56 &  & 1 & 1 &  & 1 & 3 & 6 & 50 &  &  \\
    44 & 1709E-1112N\_6838\_3742 & 44 &  & 1 & 1 & 1 &  & 3 & 6 & 50 &  &  \\
    45 & 0683E-1006N\_2732\_4164 & 58 &  &  & 1 & 1 &  & 2 & 5 & 40 &  &  \\
    \end{tabular}
\end{table}

\section{HighRes-net Architecture}
\begin{table}[!h]
    \caption{
        \texttt{HighRes-net} architecture: The ENCODE module converts each input low-res revisit into an encoding, and it inputs 6 channels, that is, 3 channels (RGB) per input image (low-res revisit + reference frame concatenated in the channels dimension). The FUSE module is an operator that is applied recursively of a pair of encodings, until one encoding remains. An encoding can be either the output of the ENCODE or the FUSE module. It inputs 64 channels per input encoding, that is, 128 channels for a pair of encodings concatenated on the channels dimension. The DECODE module is a learned upsampling operator (contrary to the non-learned bilinear or bicubic upsampling), through a transpose convolution layer that outputs 3 channels (RGB). Note that the ConvTranspose2d stride also decides the upsampling factor, which must be an integer, hence an optional Upsample layer can further upscale the super-resolved image by a fractional factor, if needed.)
    }
    \label{tab:highresnet}
    \vskip 0.10in
    \begin{center}
    \begin{small}
    \begin{tabular}{rlr}
    \textsc{Module} & \textsc{Layers} & \textsc{parameters} \\
    \midrule
    \textsc{encode} & \texttt{Conv2d(in=6, out=64, k=3, s=1, p=1)} & \textsc{1216} \\
     & \texttt{PReLU} & \textsc{1} \\
     & \texttt{ResidualBlock(64)} & \textsc{73,858} \\
     & \texttt{ResidualBlock(64)} & \textsc{73,858} \\
     & \texttt{Conv2d(in=64, out=64, k=3, s=1, p=1)} & \textsc{36,928} \\
    \\
    \textsc{fuse} & \texttt{ResidualBlock(128)} & \textsc{295,170} \\
     & \texttt{Conv2d(in=128, out=64, k=3, s=1, p=1)} & \textsc{73,792} \\
     & \texttt{PReLU} & 1 \\
    \\
    \textsc{decode} & \texttt{ConvTranspose2d(in=64, out=64, k=3, s=1)} & \textsc{36,928} \\
     & \texttt{PreLU} & 1 \\
     & \texttt{Conv2d(in=64, out=3, k=1, s=1)} & \textsc{65} \\
    \\
    \textsc{residual} \\ \textsc{(optional)} & \texttt{Upsample(scale\_factor=3.0, mode=`bicubic')} & \textsc{0} \\
    \\
     & & \textsc{591,818 (total)} \\
    \end{tabular}
    \end{small}
    \end{center}
    \vskip 0.10in
\end{table}
\newpage
\section{Hyper-parameters}

\subsection{Hyper-Parameters  of SRResNet}\label{ap:hypersrrresnet}
We list a number of the other hyper-parameters used in training. We did not perform a comprehensive hyper-parameter sweep.

\begin{enumerate}
    \item Optimiser: Adam
    \item Learning Rate: 0.0007
    \item Learning Rate Decay: The learning rate was reduced when the loss on the validation set plateaued for 2 epochs.
    \item Epochs: 50
\end{enumerate}

\subsection{Hyper-Parameters of HRNet}\label{ap:hyperhr}

We list here a number of the hyper-parameters used for training the HRNet~\cite{hrnet}. We did not perform a comprehensive hyper-parameter sweep.

\begin{enumerate}
    \item Objective Function: Binary Cross Entropy
    \item Optimizer: Adam
    \item Learning Rate: 0.0007
    \item Learning Rate Decay: The learning rate was reduced when the loss on the validation set plateaued for 2 epochs.
    \item Epochs: 50
\end{enumerate}
\reftitle{References}


\newpage
\externalbibliography{yes}
\bibliography{Bibliography/RS_SR_GANs,Bibliography/references}

\end{document}